\documentclass{optica-article}

\usepackage{mathtools, amsmath, braket}
\journal{opticajournal}

\articletype{Research Article}

\begin{document}

\title{Coincidence detection for photon triplet sources}

\author{Zijun Chen\authormark{*} and Yeshaiahu Fainman}

\address{Department of Electrical and Computer Engineering, University of California at San Diego, La Jolla, California 92093-0407, USA}

\email{\authormark{*}zic005@ucsd.edu} 


\begin{abstract}
Photon triplet generation based on third-order spontaneous parametric down-conversion remains as an experimental challenge. The challenge stems from the trade-offs between source brightness and instrument noise. This work presents a probability theory of coincidence detection to address the detection limit in source characterization. We use Bayes' theorem to model instruments as a noisy communication channel and apply statistical inference to identify the minimum detectable coincidence rate. A triplet generation rate of 1--100 Hz is required for source characterization performed over 1--72 hours using superconducting nanowire single-photon detectors.  
\end{abstract}

\section{Introduction}
Third-order spontaneous parametric down-conversion is a third-order nonlinear optical process that splits the photon energy of an intense pump into a triplet of lower energy photons. The photon energies of the triplet can be any positive values that satisfy conservation of energy and momentum. This feature of parametric conversion results an inseparable joint spectrum responsible for frequency entanglement. Each frequency conversion is an event that spontaneously occurs at an uncertain time, and yet the conversion is a quasi-instantaneous event resulting a triplet of photons with a nearly identical birth time. Therefore, in source characterization, we leverage interval analyzers to time photon arrivals and search for coincidences that all detection channels simultaneously registered a count. The challenge is to time rare arrivals using instruments with finite dark counts, detection efficiency, dead time and timing jitter. 

To combat the challenges in detection, researchers aim to increase the triplet generation rate by engineering nonlinear optical materials and devices. Early studies \cite{2004Douady, 2008Gravier} acknowledged the challenge in observing triplets. Instead of tackling third-order spontaneous parametric down-conversion directly, the researchers demonstrated third-order difference frequency generation in potassium titanyl phosphate crystals. The seeding proliferates the difference frequency to a detectable intensity and provides an alternative path to progress on technology development. The efforts resulted the making of potassium titanyl phosphate waveguides \cite{2022Bencheikh} to further improve efficiency. Potassium titanyl phosphate, however, is susceptible to competitions between second-order and third-order nonlinear optical processes \cite{2004Douady}. Such a challenge can be overcome using centrosymmetric materials with vanishing second-order nonlinearity. Researchers considered silica \cite{2011Corona_a, 2011Corona_b, 2016Cavanna}, titanium dioxide \cite{2016Moebius} and silicon nitride \cite{2016Akbari, 2022Banic} for third-order spontaneous parametric down-conversion thanks to their transparency range and manufacturability. In these works, the common trend is to optimize the triplet generation rate but proceeds to demonstrate third-harmonic generation in experiments. The third-harmonic conversion efficiency is a means to infer the triplet generation rate. Direct observation of third-order spontaneous parametric down-conversion is yet to be demonstrated in optics.

The approaches used in source characterization desire a method to address the detection limit in coincidence counting. Indeed Borshchevskaya \textit{et al.} \cite{2015Borshchevskaya} recognized the problem and assessed the integration time required to detect third-order spontaneous parametric down-conversion from calcite and rutile crystals. The study determined the required integration time from 0 dB signal-to-noise ratio for a given set of generation rate, detection efficiency, dark count rate and timing jitter. The missing element is a systemic method to derive the results presented in Ref. \cite{2015Borshchevskaya}. Such a method is invaluable for experimenters who want to assess their own systems. In addition, the integration time is typically limited to at most 100 hours due to finite memory and communication speed. Knowing the generation rate sufficient for short interval acquisition is vital.

Here we present a probability theory of coincidence detection to address the detection limit in source characterization. Section \ref{sec: method} presents the methods to assess the generation rate and integration time required for source characterization. The methods consider an $n$-fold coincidence detection system as a noisy communication channel and apply Bayes' theorem to examine the detection limit in coincidence counting. Section \ref{sec: cdet} uses combinatorial logics to model instrument noise. The results reveal how noisy instruments impact the trade-offs between source brightness, integration time and coincidence-to-accidental ratio. Section \ref{sec: conclusion} presents our conclusion and remarks. Our approach is adaptive and applicable to $n$-fold coincidence detection systems.

\section{Methods}\label{sec: method}
\begin{figure}[ht!]
\centering
\includegraphics[width = 5.0 in]{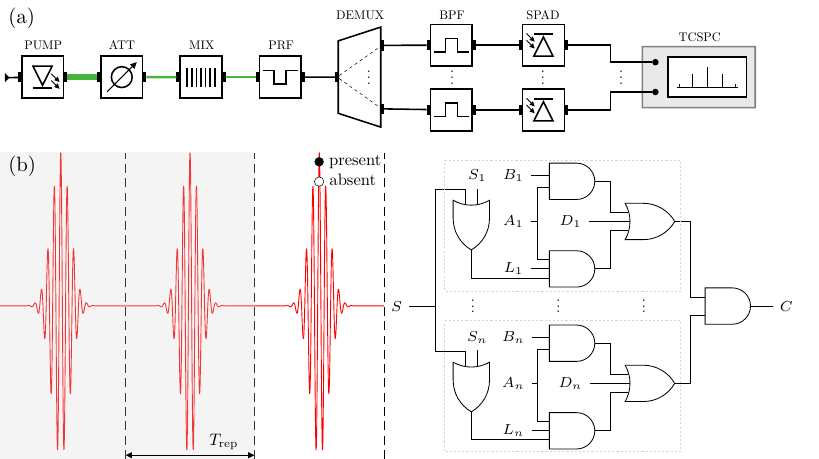}
\caption{
(a) Block diagram and (b) Boolean logic of an $n$-fold coincidence detection system. PUMP: pump laser; ATT: attenuator; MIX: optical mixer; PRF: pump-rejection filter; DEMUX: demultiplexer; BPF: bandpass filter; SPAD: single-photon avalanche diode; TCSPC: time-correlated single photon counting module; $T_{\text{rep}}$: repetition period; $A_{\nu}$: avalanche (detection efficiency); $B_{\nu}$: background noise; $C$: coincidence; $D_{\nu}$: dark count; $L_{\nu}$: path loss; $S_{\nu}$: signal generation noise; $S$: signal.     
}
\label{fig1: system}
\end{figure}

\textbf{Overview.} We want to assess the generation rate and integration time required for source characterization. The assessment needs a method to evaluate the system performance that depends on the generation rate, dark counts, detection efficiency and dead time. The approach here considers an $n$-fold coincidence detection system shown in Figure \ref{fig1: system} as a noisy communication channel. We use Bayes' theorem to link well-known methods in quantum optics and information theory and apply statistical inference to identify the detection limit in coincidence counting.

\textbf{Generation rate.} The generation rate of a time-correlated photon source under pulsed pumping is given by the number of coincidences recorded by noiseless instruments in one excitation cycle. To predict the generation rate,  we use the correspondence rule to identify the Hamiltonian \cite{2020DominguezSerna} from Poynting's theorem and apply time-dependent perturbation theory \cite{Sakurai:2021:0} to solve the Schr{\"o}dinger equation up to first-order accuracy. Time-dependent perturbation theory assumes the free-field Hamiltonian has a known set of eigenstates. The eigenstates are multimode number states and are determined from solving the time-independent Schr{\"o}dinger equation. We use the single-mode single-photon states to construct a quantum operator of coincidence detection. The coincidence detector is an $n$-fold tensor product of single-photon projectors, with $n$ being the number of detection channels. Each channel has a detection band modeled as a Gaussian spectral filter \cite{2023ZChen_coherence}. Knowing the quantum operator and state, Born's rule is applied to evaluate the generation rate of a grated waveguide under the approximations that (\romannumeral1) the modes are copolarized and (\romannumeral2) each mode experiences negligible dispersion within its detection band. Then the triplet generation rate $R^{(3)}$ is given by
\begin{equation}
R^{(3)}
	\equiv
	\frac{p_{S}}{T_{\text{rep}}} 
	\approx
	\frac{\pi^{3}\Delta f_{1}\Delta f_{2} \Delta f_{3}}{\ln(2)\Delta f_{\text{p}}}
	\left[
	1
	+
	\sum_{\nu = 1}^{3}\left( \frac{\Delta f_{\nu}}{\Delta f_{\text{p}}} \right)^{2}
	\right]^{-1/2}
	\frac{\hbar\Omega_{1}\hbar\Omega_{2}\hbar\Omega_{3}}{2^{5}\pi^{2} (\hbar\Omega_{\text{p}})^{2}}
	\left(
	\Delta\gamma_{\text{eff}}
	L_{\text{eff}}
	\right)^{2}
	P_{\text{p}}
	,
\end{equation}
where $1 / T_{\text{rep}}$ is the repetition rate of a pump laser, $p_{S}$ is the arrival probability of the triplet signal, $\Delta \gamma_{\text{eff}}$ is the effective nonlinearity, $L_{\text{eff}}$ is the effective interaction length, $P_{\text{p}}$ is the average power of the pump, $\hbar\Omega_{\mu}$ is the photon energy, $\Delta f_{\mu}$ is the full-width-at-half-maximum (FWHM) bandwidth, and $\mu = S, \text{p}, 1, 2, 3$ abbreviate the triplet signal, pump and detection channel numbers. 

\begin{figure}[ht!]
\centering
\includegraphics[width = 5.0 in]{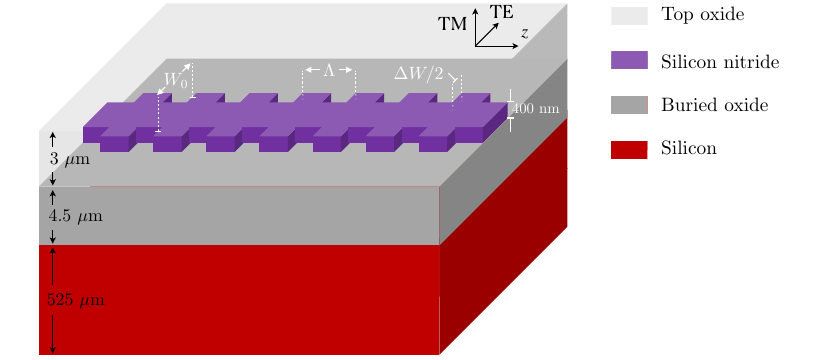}
\caption{Photon triplet source based on silicon nitride. $\Lambda$: grating period; $W_{0}$: central width; $\Delta W / 2$: modulation depth. }
\label{fig:device}
\end{figure}

\begin{table}[ht!]
	\begin{tabular}{clccccc}
		\hline\hline
		Symbol				&						& Ch. 1 	& Ch. 2 	& Ch. 3 	& Pump 	& Unit	\\
		\hline
		$\lambda_{\nu}$		& center wavelength			& 1522	& 1560 	& 1600 		& 520 	& nm		\\
		$\hbar\Omega_{\nu}$	& photon energy			& 0.815	& 0.795	& 0.775 	 	& 2.38 	& eV		\\
		$\Delta\lambda_{\nu}$	& bandwidth (FWHM	)		& 12	 	& 12 		& 12 			& $7.96\times 10^{-5}$ 	& nm		\\
		$\Delta f_{\nu}$			& bandwidth (FWHM)		& 1.55 	& 1.48	& 1.41 		& $8.83\times 10^{-5}$ 	& THz	\\
		$p_{A_{\nu}}$			& detection efficiency 		& 0.9		& 0.9		& 0.9 		& -		& -		\\
		$p_{B_{\nu}}$			& background noise			& 0		& 0		& 0 			& -		& -		\\
		$p_{D_{\nu}}$			& dark count				& 5		& 5		& 5			& -		& $\times10^{-6}$		\\
		$p_{L_{\nu}}$			& path loss				& 0.1		& 0.1		& 0.1			& -		& -		\\
		\multicolumn{1}{c}{$\Delta\gamma_{\text{eff}}$} 	& \multicolumn{1}{l}{effective nonlinearity}	&  \multicolumn{4}{c}{0.234} &  \multicolumn{1}{c}{$\text{W}^{-1}\text{m}^{-1}$}	\\
		\multicolumn{1}{c}{$L_{\text{eff}}$} 	& \multicolumn{1}{l}{interaction length}	&  \multicolumn{4}{c}{9.8} &  \multicolumn{1}{c}{mm}	\\
		\multicolumn{1}{c}{$W_{0}$} 		& \multicolumn{1}{l}{central width}		&  \multicolumn{4}{c}{1350} &  \multicolumn{1}{c}{nm}	\\
		\multicolumn{1}{c}{$\Delta W / 2$}	& \multicolumn{1}{l}{modulation depth}	&  \multicolumn{4}{c}{150} &  \multicolumn{1}{c}{nm}	\\
		\multicolumn{1}{c}{$\Lambda$} 		& \multicolumn{1}{l}{grating period}	 	&  \multicolumn{4}{c}{1701} &  \multicolumn{1}{c}{nm}	\\
		\multicolumn{1}{c}{$P_{\text{p}}$} 	& \multicolumn{1}{l}{pump power}	&  \multicolumn{4}{c}{< 12} &  \multicolumn{1}{c}{mW}	\\
		\multicolumn{1}{c}{$1/T_{\text{rep}}$} 	& \multicolumn{1}{l}{repetition rate}		&  \multicolumn{4}{c}{10} &  \multicolumn{1}{c}{MHz}		\\
		\hline\hline
	\end{tabular}
	\centering
	\caption{System and device parameters. $p_{\Box}$: occurrence probability of $\Box$.}
	\label{table: param}
\end{table}

\textbf{Device.} Figure \ref{fig:device} presents a grated waveguide for generating photon triplets on silicon nitride chips. The geometry is optimized for the triplet generation rate under constraints imposed by the pump laser, detectors and fabrication. Table \ref{table: param} summarizes the parameters being considered. The pump is a nanosecond pulsed laser system that supports a pulse duration of 5 ns at a center wavelength of 520 nm and a repetition rate of 10 MHz. The pump is edge coupled to the fundamental quasi-transverse electric (quasi-TE$_{00}$) mode of an inverse taper and adiabatically delivers energy into a grated waveguide. The grated waveguide is a parametric mixer that converts the pump energy into a triplet of photons occupying the quasi-TE$_{00}$ mode at 1522, 1560 and 1600 nm center wavelengths. We consider fabricating the device on a silicon nitride platform offered by multiproject wafer services. The technology \cite{ANT_2023} consists of material layers and thicknesses shown in Figure \ref{fig:device}. The fabless approach constraints design strategies to sizing the geometry. The waveguide length is chosen at 1.55 cm and corresponds to a loss-limited interaction length of 9.8 mm \cite{2023ZChen_THG}, assuming a loss of 1.4 dB/cm \cite{ANT_2023} for all modes. The interaction length is extended by quasi-phase matching. The sidewall modulation results an effective nonlinearity that is 10 times smaller than that of the core. To improve in the effective nonlinearity \cite{2023ZChen_THG}, a central width of 1350 nm and a modulation depth of 150 nm are picked. This design choice trades in the single-mode operation at telecom wavelengths. The device can achieve a generation rate of 0.13 mHz when pumped at 10 dBm.

\textbf{Noise.} The generation rate modeling is commonly used to optimize source brightness. The method, however, excludes noise in detectors. An additional probabilistic model is needed to capture how noisy instruments distort the generation rate in source characterization. The approach here considers coincidences as a sequence of independent and identically distributed Bernoulli random variables. Each random variable denotes two possible outcomes observed in one excitation cycle. Then, according to Bayes' theorem, the relation between the signal $S$ and the coincidence $C$ is given by the transformation
\begin{equation}
\label{eq:pc=T.ps}
\begin{bmatrix}
	p_{\overline{C}}\\
	p_{C}
\end{bmatrix}
=
\begin{bmatrix}
	1 - p_{C \mid \overline{S}}		& 1 - p_{C \mid S}	\\
	p_{C \mid \overline{S}} 		& p_{C \mid S}
\end{bmatrix}
\begin{bmatrix}
	p_{\overline{S}}\\
	p_{S}
\end{bmatrix}
\equiv
\mathbb{T}
\begin{bmatrix}
	p_{\overline{S}}\\
	p_{S}
\end{bmatrix},
\end{equation}
with the arrival probability $p_{S}$ and the coincidence probability $p_{C}$ satisfying
\begin{subequations}
\begin{alignat}{2}
& p_{\overline{S}}
	\equiv \Pr(S = 0)
	= 1 - p_{S}
	,
	\qquad
&& p_{S}
	\equiv \Pr(S = 1)
	= R^{(n)}T_{\text{rep}}
	,
	\\
& p_{\overline{C}}
	\equiv \Pr(C = 0)
	= 1 - p_{C}
	,
	\qquad
&& p_{C}
	\equiv \Pr(C = 1)
	= R_{C}T_{\text{rep}},
\end{alignat}
\end{subequations}
where $\Pr(\cdot)$ is the probability of an event, $T_{\text{rep}}$ is the repetition period, $R_{C}$ is the coincidence rate, and $R^{(n)}$ is the generation rate for $n$th-order spontaneous parametric down-conversion. The elements of the transfer matrix $\mathbb{T}$ are conditional probability given by
\begin{subequations}
\begin{equation}
\label{eq:Pr(C=1|S=0)}
p_{C \mid \overline{S}}
	\equiv \Pr(C = 1 \mid S = 0)
	= \Pr\{C(S = 0) = 1\}
	,
\end{equation}
\begin{equation}
\label{eq:Pr(C=1|S=1)}
p_{C \mid S}
	\equiv \Pr(C = 1 \mid S = 1)
	= \Pr\{C(S = 1) = 1\}
	.
\end{equation}
\end{subequations}
The transfer matrix characterizes the noise performance of a coincidence detection system. The abstraction is useful when the transfer matrix elements are expressed in instrument specifications. We implement coincidence logics to model instrument noise and derive from Eq. \eqref{eq:Pr(C=1|S=0)} and \eqref{eq:Pr(C=1|S=1)} an explicit relation between the signal, coincidence and noise. Section \ref{sec: cdet} presents the transfer matrix elements for two coincidence logics with increasing complexity. The transfer matrix elements are derived assuming all noise sources are independent.  

\textbf{Channel capacity.} An ideal detection system replicates the arrival information of a signal. Noise, however, sets the maximum capacity that a system can preserve arrival information. The channel capacity \cite{1948Shannon_communication} is found from maximizing the mutual information shared between the signal and coincidence. Assuming the transfer matrix is independent of the arrival probability, the channel capacity $\mathfrak{C}$ is
\begin{equation}
\mathfrak{C}
= \log_{2}\left\{ 1 + 2^{\displaystyle-[H_{b}(p_{C \mid S}) - H_{b}(p_{C \mid \overline{S}})] / \det(\mathbb{T})} \right\}
+ \frac{ p_{C \mid \overline{S}} H_{b}(p_{C \mid S}) - p_{C \mid S} H_{b}(p_{C \mid \overline{S}})}{\det(\mathbb{T})},
\end{equation}
with $H_{b}(p) = -p\log_{2}(p) - (1-p)\log_{2}(1-p)$ being the binary entropy function.

\textbf{Inference.} The purpose for recording coincidences is to infer the arrivals of a signal. The inference is made with uncertainty given the data are acquired through noisy instruments. To judge if a coincidence reliably announces an arrival, a measurement-based inference should be more reliable than a random guess. This requirement can be written as
\begin{equation}
\label{eq: binaryInference}
p_{S \mid C} = \frac{p_{C \mid S}p_{S}}{p_{C}} > 50\%.
\end{equation}
Equation \eqref{eq: binaryInference} is applied as an inference criterion to identify the minimum detectable coincidence rate. 
\section{Results} \label{sec: cdet}
Section \ref{sec: method} presented the methods for assessing the source brightness and instrument noise. To quantify the performance of a coincidence detection system, the transfer matrix elements should be related to instrument specifications. In this section we use combinatorial logic gates to model noisy coincidence detection systems and to examine the minimum detectable rate and integration time required for source characterization.

\subsection{Coincidence logics}
\begin{figure}[ht!]
\centering
\includegraphics[width = 5.0 in]{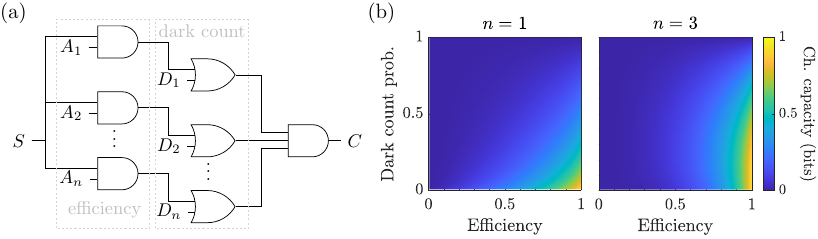}
\caption{ (a) Boolean logics and (b) channel capacity of $n$-fold coincidence. $A_{\nu}$: avalanche (detection efficiency); $C$: coincidence; $D_{\nu}$: dark count; $S$: signal.}
\label{fig: det}
\end{figure}

Figure \ref{fig: det}(a) illustrates the main features in coincidence logics. The input stage models the arrival of time-correlated photons. We consider time-correlated photons as a set of down-converted photons having the same arrival time \cite{1970Burnham_simultaneity}. The signal therefore lights up all channels at the same time. The logic gates in each channel model the detection efficiency and dark counts. The AND operation between the signal $S$ and the avalanche $A_{\nu}$ has the following implication. In the absence of a dark count $D_{\nu}$, a singles count marks an arrival only if the arrival has been registered. The probability of success in registering an arrival is the detection efficiency. In the presence of a dark count, the OR gate refines the logic to that a singles count either marks a registered arrival or a dark count. The last stage is an $n$-input AND gate implementing the $n$-fold coincidence logic. A coincidence count is registered only when all channels have registered singles counts within the coincidence time window. 

Modeling coincidence detection as logic gates is useful in assessing system performance. Consider the coincidence $C$ as a Boolean expression
\begin{equation}
C = (A_{1}S + D_{1})(A_{2}S + D_{2})\dots(A_{n}S + D_{n}),
\end{equation}
then the transfer matrix elements are found to be
\begin{subequations}
\begin{align}
p_{C \mid S}
	& = \prod_{\nu = 1}^{n} \left(p_{\overline{D}_{\nu}}p_{A_{\nu}} + p_{D_{\nu}} \right),
	\\
\label{eq: pD1pD2...pDn}
p_{C\mid\overline{S}} 
	& = p_{D_{1}}p_{D_{2}} \dots p_{D_{n}}.
\end{align}
\end{subequations}
Equation \eqref{eq: pD1pD2...pDn} states the probability of false coincidence is an $n$-fold product of the dark count probabilities. This implies multichannel coincidence detection suppresses false coincidences at the expense of increasing multiplicative noise. The trade-off can be inferred from the determinant
\begin{equation}
\det(\mathbb{T})
=
\begin{dcases}
1 - p_{D_{1}}p_{D_{2}} \dots p_{D_{n}},	& \text{if } p_{A_{1}}, p_{A_{2}}, \dots, p_{A_{n}} = 1,	\\
p_{A_{1}}p_{A_{2}} \dots p_{A_{n}},		& \text{if } p_{D_{1}}, p_{D_{2}}, \dots, p_{D_{n}} = 0.	\\
\end{dcases}
\end{equation}
Figure \ref{fig: det}(b) further demonstrates the trade-off between the detection efficiency and dark counts. We conclude that efficient detectors enlarge the noise margin of dark counts in a multichannel coincidence detection system. The noise equivalent power of one channel cannot conclusively assess the minimum detectable coincidence rate.

Registration of a coincidence takes time. The time required to resolve two consecutive arrivals is the dead time \cite{Knoll:2010:ch4.7deadtime}. Dead time results lost coincidences in data acquisition. An AND gate can be used to model the dead-time loss in coincidence logics. Let raw count rates be the count rates that are uncorrected for the dead-time loss and suppose the product of the raw singles rate $R_{Z_{\nu}}$ and the dead time $T_{\text{dt}_{\nu}}$ amounts the probability that the detector in the $\nu$th channel is inactive, then the raw coincidence rate $R_{Z}$ is
\begin{equation}
R_{Z} 
= R_{C}\prod_{\nu = 1}^{n} \left(1 - R_{Z_{\nu}}T_{\text{dt}_{\nu}} \right).
\end{equation}
Note the raw count rates are quantities directly acquired from the readouts of a time-correlated single-photon counting module using the expression \cite{Beck:2007:rate-count-probability, 2013Stevens_ch2}
\begin{equation}
\label{eq:Rz}
R_{Z_{X}}
	= \frac{N_{Z_{X}}}{T_{\text{int}}}
	= \frac{p_{Z_{X}}}{T_{\text{rep}}}
	,
\end{equation}
where $Z_{X}$ is the Bernoulli random variable of $X$ that is uncorrected for the dead-time loss and $N_{Z_{X}}$ is the number of events recorded over an integration time $T_{\text{int}}$. Then the generation rate can be expressed as
\begin{equation}
\label{eq:R(n)primitive}
R^{(n)} 
= 
\frac{		R_{Z} - R_{C \mid \overline{S} } \prod_{\nu = 1}^{n}\left(1 - R_{Z_{\nu}}T_{\text{dt}_{\nu}}\right)}
	{	\det(\mathbb{T})\prod_{\nu = 1}^{n}\left(1 - R_{Z_{\nu}}T_{\text{dt}_{\nu}}\right)}
\equiv
\frac{		R_{Z} - R_{Z \mid \overline{S} }}
	{	\det(\mathbb{T})\prod_{\nu = 1}^{n}\left(1 - R_{Z_{\nu}}T_{\text{dt}_{\nu}}\right)}
	.
\end{equation}
The challenge in applying Eq. \eqref{eq:R(n)primitive} is the need to measure the transfer matrix elements. In addition, the false coincidence rate $R_{Z \mid \overline{S}}$ is equal to the dark coincidence rate measured in the absence of a signal. This equality holds true only when the source is noiseless. The assumption cannot be made since any realistic source has noise.

\subsection{Source characterization} \label{sec: sourceChar}
Coincidence logics need to account for noise associated with time-correlated photon sources. Let the coincidence logics shown in Figure \ref{fig1: system}(b) be a Boolean expression of the form
\begin{subequations}
	\begin{align}
C 		& = C_{1}C_{2} \dots C_{n},		\\
C_{\nu} 	& = A_{\nu}\left[L_{\nu}\left(S + S_{\nu}\right) + B_{\nu}\right] + D_{\nu} \; 
\quad (\nu = 1, 2, \dots, n),
\end{align}
\end{subequations}
then the transfer matrix elements are found to be
\begin{subequations}
\begin{align}
p_{C \mid S}
&= \prod_{\nu = 1}^{n}\left[p_{\overline{D}_{\nu}}p_{A_{\nu}}\left(p_{\overline{B}_{\nu}}p_{L_{\nu}} + p_{B_{\nu}} \right) + p_{D_{\nu}} \right], \\
p_{C \mid \overline{S}}
\label{pC|Sbar_advanced}
&= \prod_{\nu = 1}^{n}\left[ p_{\overline{D}_{\nu}}p_{A_{\nu}}\left(p_{\overline{B}_{\nu}}p_{L_{\nu}}p_{{S}_{\nu}} + p_{B_{\nu}}\right) + p_{D_{\nu}}\right].
\end{align}
\end{subequations}
The Bernoulli random variables denote the following: $A_{\nu}$ is the avalanche (detection efficiency); $B_{\nu}$ is the background noise; $C_{\nu}$ is the singles; $C$ is the coincidence; $D_{\nu}$ is the dark count; $L_{\nu}$ is the path loss; $S$ is the signal; $S_{\nu}$ is the signal generation noise. The signal generation noise appearing in each channel are independent and identically distributed random variables such that $p_{S_{\nu}} = p_{S}$ for $\nu = 1, 2, \dots, n$. The signal generation noise models a subset of time-correlated photons whose arrival time are decorrelated by unknown causes.

The relation between the generation rate and raw data was demonstrated at Eq. \eqref{eq:R(n)primitive}. The challenge is the false coincidence rate cannot be extracted from measuring the dark coincidence rate. To mark the distinction, the dark coincidence probability $p_{\text{dark}}$ is defined as the probability of measuring a coincidence when the source is off,
\begin{equation}
\label{eq:pdark}
p_{\text{dark}} 
\equiv
\left.p_{C}\right\rvert_{p_{S} = 0}
= \prod_{\nu = 1}^{n}\left( p_{\overline{D}_{\nu}}p_{A_{\nu}}p_{B_{\nu}} + p_{D_{\nu}}\right).
\end{equation}
Comparison of Eq. \eqref{pC|Sbar_advanced} and \eqref{eq:pdark} shows that the probability of false coincidence coincides with the probability of dark coincidence only when the signal generation noise is absent. Let us further define the probability $p_{\text{acc}}$ of accidental coincidence as
\begin{equation}
\label{eq: pacc}
p_{\text{acc}} 
= p_{C} - \prod_{\nu = 1}^{n}p_{L_{\nu}}p_{A_{\nu}} p_{S}
.
\end{equation}
The accidental coincidence is the most general form of noise. An accidental coincidence is any coincidence that deviates from the expected signal. The accidental rate is useful in relating the generation rate to observables. The insight roots in the attribute of time-correlated photons. The signal's arrival is unique in lighting up all channels at the same time. The signature distinguishes the arrival time of the signal from that of noise. The raw accidental rate $R_{Z_{\text{acc}}}$ can therefore be acquired from the readouts recorded in the nonzero delay bins of the coincidence histogram. Consider an alternative form of the generation rate
\begin{equation}
\label{eq:R(n)advanced}
R^{(n)} = 
\frac{R_{Z} - R_{\text{acc}} \prod_{\nu = 1}^{n}\left(1 - R_{Z_{\nu}}T_{\text{dt}_{\nu}}\right)}{\prod_{\nu = 1}^{n} p_{L_{\nu}}p_{A_{\nu}}\left(1 - R_{Z_{\nu}}T_{\text{dt}_{\nu}}\right)}
\equiv
\frac{R_{Z} - R_{Z_{\text{acc}}}}{\prod_{\nu = 1}^{n} p_{L_{\nu}}p_{A_{\nu}}\left(1 - R_{Z_{\nu}}T_{\text{dt}_{\nu}}\right)}.
\end{equation}
Equation \eqref{eq:R(n)advanced} suggests the following procedure to characterize the generation rate: (\romannumeral1) equalize the optical path length of all detection channels, (\romannumeral2) turn on the source and use the time-correlated single photon counting module in time-tag mode to acquire the singles counts and coincidences, (\romannumeral3) turn off the source and use the time-correlated single photon counting module in time-tag mode to measure dark coincidences and (\romannumeral4) characterize the path loss, detection efficiency and dead time of each channel. Then the generation rate can be evaluated from the acquired data according to Eq. \eqref{eq:R(n)advanced}. 

In addition to source brightness, the noisiness of a source is a crucial metric to evaluate device performance. The noise performance of a time-correlated photon source is quantified by the coincidence-to-accidental ratio (CAR) \cite{2012Davanco} defined as the ratio of the expected signal to the source-induced noise,
\begin{equation}
\label{eq: CAR}
\text{CAR} 
	=
	\frac{p_{C} - p_{\text{acc}}}{p_{\text{acc}} - p_{\text{dark}} }
	=
	  \frac{R_{Z} - R_{Z_{\text{acc}}}}
	 {R_{Z_{\text{acc}}} - R_{Z_{\text{dark}}}}
	 .
\end{equation}
Suppose the coincidence detection system has no additive noise such that $p_{B_{\nu}} = 0$ and $p_{D_{\nu}} =0$ for $\nu = 1, 2, \dots, n$, then
\begin{equation}
\text{CAR}
	=
	\frac{p_{S}}{p_{\overline{S}}p_{{S}_{1}}p_{{S}_{2}} \dots p_{{S}_{n}}}
	.
\end{equation}
The coincidence-to-accidental ratio reduces to the signal-to-noise ratio intrinsic to the source. The ratio of the expected signal to the source-induced noise provides the best estimate on the noise performance of a time-correlated photon source. There is no apparent technique to directly probe the intrinsic signal-to-noise ratio in experiments without making assumptions.

\begin{figure}[ht!]
\centering
\includegraphics[width = 5.0 in]{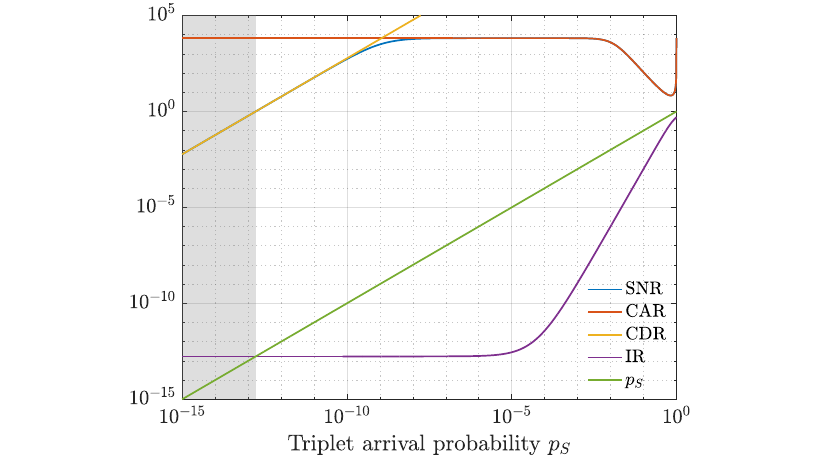}
\caption{Minimum arrival probability in a three-fold coincidence detection system. SNR: signal-to-noise ratio; CAR: coincidence-to-accidental ratio; CDR: coincidence-to-dark ratio; IR: inference ratio.}
\label{fig: minDetectableCoincidence}
\end{figure}

\begin{figure}[ht!]
\centering
\includegraphics[width = 5.0 in]{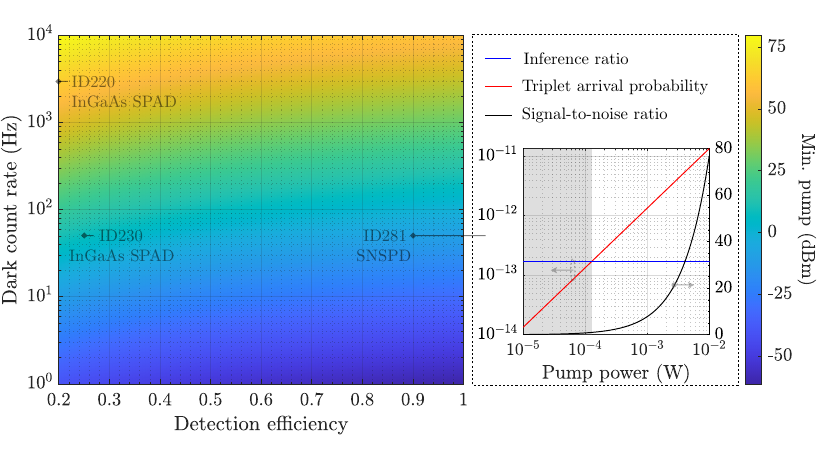}
\caption{Design space of detectors and triplet generators.}
\label{fig: minPumpPower}
\end{figure}

Any detection system has a minimum detectable rate. In radiometry the minimum detectable power is the power that yields a signal-to-noise ratio of one. Here we define the signal-to-noise ratio (SNR) as
\begin{equation}
\text{SNR} =
	\frac{p_{C} - p_{\text{acc}}}{p_{\text{acc}}}
	=
	\frac{R_{Z} - R_{Z_{\text{acc}}}}{R_{Z_{\text{acc}}}}
	.
\end{equation}
The difference between the signal-to-noise ratio and coincidence-to-accidental ratio is best illustrated by introducing the coincidence-to-dark ratio (CDR) of the form
\begin{equation}
\text{CDR} = 
	\frac{p_{\text{C}} - p_{\text{acc}}}{p_{\text{dark}}}
	=
	 \frac{R_{Z} - R_{Z_{\text{acc}}}}
	 {	R_{Z_{\text{dark}}}}
	 .
\end{equation}
Figure \ref{fig: minDetectableCoincidence} compares the signal-to-noise ratio, coincidence-to-accidental ratio and coincidence-to-dark ratio. The signal-to-noise ratio approaches to the coincidence-to-accidental ratio when the measurement is limited by the signal generation noise. In contrast the signal-to-noise ratio approaches to the coincidence-to-dark ratio as the arrival probability decreases. The measurement is instrument-limited when the minimum detectable rate is a concern. Setting the coincidence-to-dark ratio at one is sufficient to determine the minimum detectable rate. An alternative method to assess the minimum detectable rate is statistical inference. The purpose for recording a coincidence is to infer the signal arrival. The inference can be made with certainty if
\begin{equation}
	p_{S}
	>
	\left(	
	\frac{p_{C\mid\overline{S}}}{p_{C\mid\overline{S}} + p_{C \mid S}}
	\equiv	
	\text{IR}
	\right)
	;
\end{equation}
that is, the arrival probability should at least equal to the inference ratio (IR). The minimum arrival probability falls within the instrument-limited regime according to Figure \ref{fig: minDetectableCoincidence}. Consider
\begin{equation}
\min(p_{S})
	=
\displaystyle
\begin{dcases}
\frac{ p_{\text{dark}}}{\prod_{\nu = 1}^{n} p_{A_{\nu}}p_{L_{\nu}}},			& \text{if } \text{CDR} = 1 \text{ is used}, \\
\frac{p_{C \mid \overline{S}} }{p_{C \mid \overline{S}} + p_{C \mid S}},		& \text{if } \text{IR} = p_{S} \text{ is used} .
\end{dcases} 
\end{equation}
Both methods agree when (\romannumeral1) the signal generation noise is negligible and (\romannumeral2) $p_{C \mid S} \gg p_{C \mid \overline{S}}$. Conditions (\romannumeral1) and (\romannumeral2) hold true in practice. We conclude the minimum detectable rate is
\begin{equation}
\label{eq: Rsmin}
R_{\text{min}}^{(n)}
	=
	\frac{R_{\text{dark}}}{\prod_{\nu = 1}^{n} p_{A_{\nu}}p_{L_{\nu}}}
	.
\end{equation}
Note the minimum detectable rate is source independent. The noise performance of detectors dictates the noise floor. Figure \ref{fig: minPumpPower} shows the pump power required to yield a minimum detectable coincidence rate. The markers reference three commercial single-photon detectors \cite{IDQ_2023}. ID220 and ID230 are InGaAs single-photon avalanche photodiodes (SPAD), whereas ID281 is a superconducting nanowire single-photon detector (SNSPD). We chose ID281 because the pump laser has a maximum power of 12 mW. Using a dark count rate of 50 Hz and a detection efficiency of 90\%, a signal-to-noise ratio of 19 dB is achieved at an on-chip pump power of 10 dBm. 

The remaining challenge in source characterization is the integration time required to accumulate a sensible number of coincidences. Consider estimating from Eq. \eqref{eq:Rz} the integration time required to register 100 coincidences \cite{2010Hubel, 2012Shalm, 2014Hamel, 2017Agne}:
\begin{equation}
\label{eq: Tmin}
	\frac{T_{\text{int}}}{N_{Z}}
	\approx
	\frac{T_{\text{rep}}}{\det(\mathbb{T}) p_{S} + p_{C \mid \overline{S}}}
	.
\end{equation}
Application of Eq. \eqref{eq: Tmin} reveals a triplet generation rate of 1--100 Hz is required to accumulate 100 coincidences within 1--72 hours using a superconducting nanowire single-photon detection system. The submillihertz triplet generation rate achievable in a grated silicon nitride waveguide demands a prohibitive integration time. 

\section{Conclusion}\label{sec: conclusion}
In conclusion we presented a coincidence detection theory to examine the trade-offs between source brightness and instrument noise. To quantify noise performance, Bayes' theorem and combinatorial logics are used to model detection efficiency, dark counts, dead time and signal generation noise. Analysis revealed coincidence detection is superb in suppressing false alarms caused by dark counts but at the expense of lowering the system efficiency. The detection limit in coincidence counting consequently differs from that of radiometry. The noise equivalent power of one channel cannot conclusively assess the noise floor of a coincidence detection system. The minimum detectable coincidence rate is determined by the dark coincidence rate and system efficiency. The challenge in source characterization is to accumulate a sensible number of coincidences within 1--72 hours. We found a triplet generation rate of 1--100 Hz is required for source characterization performed over 1--72 hours using superconducting nanowire single-photon detectors. The assessment is limited to the performance achievable in the chosen technologies. The results suggest system integration with improvements on optical pumping, single-photon detection and nonlinear optical materials will ease the challenges in demonstrating third-order spontaneous parametric down-conversion on photonic integrated circuits.

\begin{backmatter}

\bmsection{Funding}
National Science Foundation (ECCS-2023730, ECCS-2217453, ECCS-2025752); Army Research Office; Cymer.

\bmsection{Acknowledgements}
This work was supported by the National Science Foundation (NSF) grants NSF ECCS-2023730 and NSF ECCS-2217453, the Army Research Office (ARO), the San Diego Nanotechnology Infrastructure (SDNI) supported by the NSF National Nanotechnology Coordinated Infrastructure (grant ECCS-2025752), and the ASML/Cymer Corporation.

\bmsection{Disclosures}
The authors declare no conflicts of interest.

\bmsection{Data availability}
Data underlying the results presented in this paper are not publicly available at this time but may be obtained from the authors upon reasonable request.

\end{backmatter}

\bibliography{coincidenceDet_ref}

\end{document}